\documentclass[graphics,floatfix, footinbib,tightenlines,nobibnotes, aps, prb, twocolumn]{revtex4-1}
\usepackage{amsmath,amssymb,wasysym}
\usepackage{graphicx}% Include figure files
\usepackage{dcolumn}% Align table columns on decimal point
\usepackage{bm}% bold math
\usepackage{braket}
\usepackage{subcaption}
\usepackage{verbatim}
\usepackage{float}
\usepackage{bbold}
\usepackage{color}
\usepackage{xcolor}
\usepackage{relsize}
\usepackage{amsthm}
\usepackage{enumerate}
\usepackage{soul,xcolor}
\usepackage[T1]{fontenc}
\usepackage{adjustbox}
\usepackage[colorlinks=true ,urlcolor=blue,urlbordercolor={0 1 1}]{hyperref}

\newcommand{\be}{\begin{equation}}
\newcommand{\ba}{\begin{align}}
\newcommand{\ee}{\end{equation}}
\newcommand{\bea}{\begin{eqnarray}}
\newcommand{\eea}{\end{eqnarray}}
\newcommand{\beq}{\begin{equation}}
\newcommand{\eeq}{\end{equation}}
\newcommand{\beqn}{\begin{eqnarray}}
\newcommand{\eeqn}{\end{eqnarray}}

\newcommand{\hb}[1]{{\color{blue}{{#1}}}} 
\usepackage{physics}
\usepackage{simpler-wick}

\renewcommand{\hat}[1]{{\widehat #1}}

\newcommand*{\yhz}[1]{\textcolor{red}{#1}}

\usepackage{bm}

\usepackage{array}
\newcolumntype{L}[1]{>{\raggedright\arraybackslash}p{#1}}
\newcolumntype{C}[1]{>{\centering\arraybackslash}p{#1}}
\newcolumntype{R}[1]{>{\raggedleft\arraybackslash}p{#1}}
\usepackage{multirow}

\allowdisplaybreaks

\begin{document}
\widetext

\title{ S-wave pairing in a two-orbital t-J model on triangular lattice: possible application to Pb$_{10-x}$Cu$_x$(PO$_4$)$_6$O }

\author{Hanbit  Oh$^1$}
\author{Ya-Hui Zhang$^2$}
\email{yzhan566@jhu.edu}
\affiliation{$^1$Department of Physics, KAIST, Daejeon, 34126, Republic of Korea}
\affiliation{$^2$William H. Miller III Department of Physics and Astronomy, Johns Hopkins University, Baltimore, Maryland, 21218, USA}

\date{\today}

\begin{abstract}
Recently room temperature superconductor was claimed in  Pb$_{10-x}$Cu$_x$(PO$_4$)$_6$O (also known as LK-99) with $x\in (0.9,1.1)$. Density functional theory (DFT) calculations suggest that the conduction electrons are from the doped Cu atoms with valence close to $d^{9}$.  Motivated by this picture, we build a two-orbital Hubbard model on triangular lattice formed by the $d_{xz}$ and $d_{yz}$ orbitals with total hole density (summed over spin and orbital) $n=1-p$.   When $p=0$, the system is in a  Mott insulator within this model. When $p>0$,  we derive a $t-J$ model and perform a self-consistent slave boson mean field calculation. Interestingly we find a s wave pairing in contrast to one-orbital t-J model which favors $d+id$ pairing. S wave pairing should be more robust to disorder and may lead to high Tc superconductor with sufficiently large value of $t$ and $J$.  However, the DFT calculations predict a very small value of $t$ and then the $T_c$ is expected to be small.  If LK99 is really a high Tc superconductor, ingredients beyond the current model is needed. We conjecture that the doped Cu atoms may distort the original lattice and form local clusters with smaller Cu -Cu distance and thus larger value of $t$ and $J$. Within these clusters we may locally apply our t-J model calculation and expect high Tc  s-wave superconductor. Then the superconducting islands couple together, which may eventually become a global superconductor, an insulator or even an anomalous metal depending on sample details.
 \end{abstract}

\maketitle

\textbf{Introduction} Recently there is report of  room temperature superconductivity in Pb$_{10-x}$Cu$_x$(PO$_4$)$_6$O, also called LK-99\cite{lee2023roomtemperature,lee2023superconductor}.  The experimental reproduction of the exciting discovery is still ongoing\cite{liu2023semiconducting,kumar2023synthesis,Shi,Chang,Abramian}.   In the theoretical side\cite{griffin2023origin,kurleto,Si,Scaffidi,Baskaran}, density functional theory (DFT) calculations suggest a picture of narrow bands formed by Cu 3d orbitals with the valence close to d$^9$\cite{griffin2023origin,kurleto,Si}, similar to the high Tc cuprates. 

In this paper, we build a two-orbital model on triangular lattice based on the Cu 3d $d_{xz}$ and $d_{yz}$ orbitals. Based on symmetry analysis, we find that three tight binding parameters are allowed with only the nearest neighbor hopping. We suggest one specific choice to fit the DFT band. To capture the strong correlations from the Cu 3d orbitals, we propose a two-orbital Hubbard model at total hole density $n=1-p$. $p=0$ is a Mott insulator and we derive a two-orbital t-J model for the small finite $p$ regime.  Then we apply the salve boson mean field theory to analyze the t-J model. Slave boson theory is known to be able to reproduce several essential properties of the superconductor in cuprates, including  the pairing symmetry and the doping dependence of  the pairing strength\cite{lee2006doping}. Based on slave boson mean field calculation, we identify a s$^{\prime}$-wave spin-singlet pairing with both intra-orbital and inter-orbital components, which is unusual given that usually t-J model calculation predicts higher angular momentum pairing.  A s-wave pairing is more robust to disorder and thus may support higher Tc at the same level of pairing strength, compared to $d$ wave pairing in cuprates.

While high Tc superconductor exists in this model with sufficiently large value of $t$ and $J$, we need a much higher value of $t$ than the DFT prediction to obtain a superconductor at the order of $100$ K. If LK-99 is indeed a high temperature (or even room temperature) superconductor, new ingredients must be included beyond the DFT calculations to greatly enhance the mobility of the electrons.  One conjecture we have is that the doped Cu atoms may distort the original lattice and form clusters with much smaller Cu-O or Cu-Cu distance. In this picture, we have regions which is locally described by our t-J model with large value of t and J.  Then we can have s-wave high Tc superconductor islands according to our calculation. These islands then need to couple together through the Josephson tunneling. Whether the system can be a global superconductor depends on details such as disorder strength and may vary a lot in different samples.

\begin{figure}[tb]
    \centering
    \includegraphics[width=\linewidth]{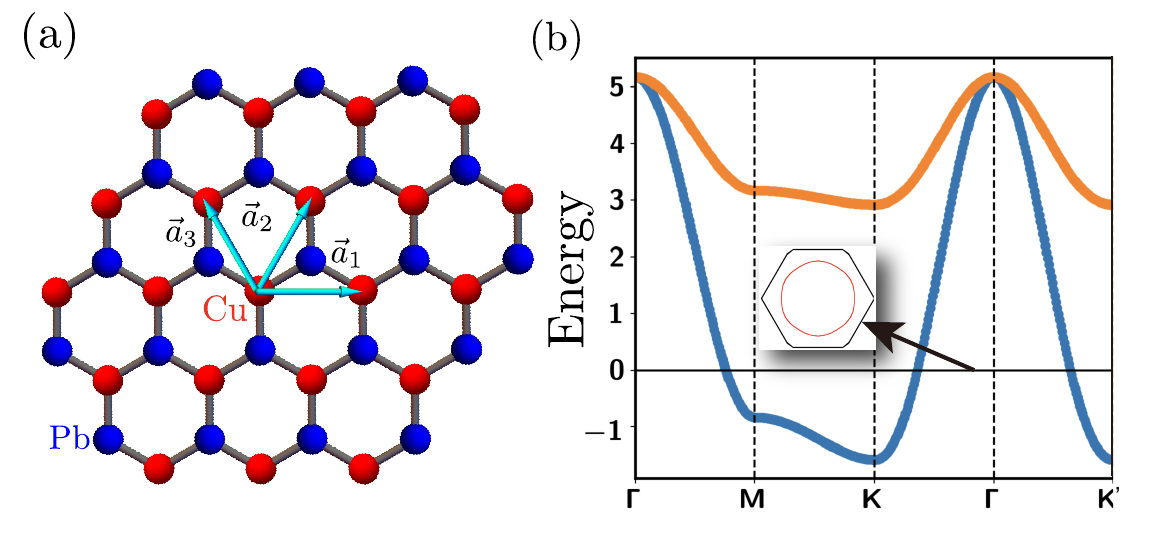}
    \caption{(a) Illustrations of atomic structure with Cu (red)and Pb (blue) atoms. The Cu itself forms a hexagonal structure with primitive vectors $\vec{a}_{i}$. Two d-orbitals ($d_{xz},d_{yz}$) are living at each Cu site. 
    (b) Energy band structure of a tight binding model (in hole picture). We set $t_{\sigma}=-1, t_{\pi}=0,t_{2} =-\sqrt{3}/4$. 
    The Fermi surfaces at total hole filling $n=1$ are illustrated in the inset. There is only a single $C_{3}$ symmetric electron-pocket near the $\Gamma$ point.}
    \label{fig:1}
\end{figure}

\textbf{Model}  We consider a two-orbital based on the $d_{xz}$ and $d_{yz}$ orbitals, living on layered triangular lattice. 
For simplicity, we will ignore the inter-layer tunneling $t_z$ in this paper and focus on a two-dimensional (2D) model on a triangular lattice, as illustrated in Fig\ref{fig:1} (a). 
 
We label $d_{i;1\sigma}$ and $d_{i;2\sigma}$ as the annihilation operator of a hole in $d_{xz}$ and $d_{yz}$ orbital respectively at site $i$ for spin $\sigma=\uparrow,\downarrow$. For convenience we use t he hole picture with the hole density $n_i=\sum_{a=1,2}\sum_{\sigma=\uparrow,\downarrow} d^\dagger_{i;a\sigma}d_{i;a\sigma}$ has an average value of $n=1-p$ per site. We will mainly consider the small positive $p$ regime. We define $\psi_{i;\sigma}=(d_{i;1\sigma},d_{i;2\sigma})^T$. Our system have a $C_3$ rotation symmetry around the site $i$. There is also a mirror reflection symmetry $\mathcal M_y$: $x\rightarrow -x, y \rightarrow y$.  Under $C_3$ and $\mathcal M_y$, we have $\psi_i \rightarrow U_{C_3} \psi_{C_3 i}$ and $\psi_i \rightarrow U_{M_y} \psi_{M_y i}$. We have $U_{C_3}=\begin{pmatrix} -\frac{1}{2}& \frac{\sqrt{3}}{2}\\ -\frac{\sqrt{3}}{2} & -\frac{1}{2}\end{pmatrix}$ and $U_{M_y}=\begin{pmatrix} -1 & 0 \\ 0 & 1 \end{pmatrix}$.  

We consider the generic form of a tight-binding model with nearest-neighbor hopping, 
\begin{eqnarray}
    H_{K}&=&
    -\sum_{\sigma=\uparrow,\downarrow}
    \sum_{i;l=1,2,3,4,5,6}  \Psi^\dagger_{i;\sigma}
\bm{T}_{l}\Psi_{i+\vec{a}_{l};\sigma} \label{eq:tightbinding}
\end{eqnarray}
with $2\times 2$ matrix of hopping matrix  $\bm{T}_{l} = \bm{T}_{-l}^{\dagger}$ as required by the Hermitian condition. Here, we used the lattice vectors which connect the nearest neighbor, $\vec{a}_{1}= (1,0),\vec{a}_{2} = (1,\sqrt{3})/2,\vec{a}_{3} = (-1,\sqrt{3})/2$. 

The hopping matrix is derived by the central symmetries of the system, $C_{3}$ rotation, and mirror symmetry $\mathcal{M}_y$. This constrains the hopping   matrix to be in the form,
\begin{equation}
    T_1=\begin{pmatrix} t_\sigma & -t_2 \\ t_2 & t_\pi \end{pmatrix},
\end{equation}
\begin{equation}
    T_2=\begin{pmatrix}
\frac {1} {4} (t_\sigma + 3 t_ {\pi}) & 
 t_ {2} + \frac {\sqrt {3}} {4} (t_\sigma- t_\pi) \\
-t_ {2} + \frac {\sqrt {3}} {4} (t_\sigma- t_\pi)
       & \frac {1} {4} (3t_\sigma +  t_ {\pi}) 
    \end{pmatrix}
\end{equation}
and 
\begin{equation}
    T_3=\begin{pmatrix}
\frac {1} {4} (t_\sigma + 3 t_ {\pi}) & - 
 t_{2} -\frac {\sqrt {3}} {4} (t_\sigma- t_\pi) \\
t_{2} - \frac {\sqrt {3}} {4} (t_\sigma- t_\pi)
       & \frac {1} {4} (3t_\sigma + t_ {\pi}) 
    \end{pmatrix}
\end{equation}
In the above $t_\sigma$ and $t_\pi$ are from the $\sigma$ and $\pi$ bond. $t_2$ arises from breaking the $C_6$ rotation symmetry. We note that  $t_2$ is necessary to split the two-fold degeneracy at K and K$^{\prime}$ point in the Brillouin zone (BZ).  While we focus on the simple 2D model, extension of it to 3D is straightforward by simply adding a term $-t_z \sum_i \Psi^\dagger_{i;\sigma} \Psi_{i+\hat{z};\sigma}$.

In the Supplemental Material (SM), we added more discussion to emphasize the symmetry action on the two orbitals.  
In Figure \ref{fig:1}(b), we illustrate the band structure model, Eq. \ref{eq:tightbinding} for a specific choice of hopping parameters:$t_\sigma=-1$, $t_\pi=0$ and $t_2=-\frac{\sqrt{3}}{4}$. Here we are using the hole picture and add a negative sign to the hopping.  In particular, at total hole filling $n=1$, there is a single electron pocket\footnote{The pocket looks like a hole pocket in the plot, but because we are using the hole picture, this is an electron pocket.} near $\Gamma$ pocket whose Fermi-surface shape is shown in the inset.  We also note that a two-orbital model was  recently already proposed\cite{Scaffidi}. But the model there seems to be not equivalent to our model at any parameter.

To also incorporate the strong on-site repulsion, we consider a Hubbard model:

\begin{equation}
    H_{Hubbard}=H_K+\frac{U}{2}\sum_i n_i (n_i-1)
\end{equation}
Note that here we ignore the difference between the intra-orbital and inter-orbital repulsion and also the Hund's coupling.

\textbf{t-J model}  When $n=1$, at large U/t regime the system is in a Mott insulator and described by a spin-orbital model at low energy.  The model can be obtained from the standard t/U expansion. To make the calculation in the next part convenient, here we represent the spin-orbital model using the Abrikosov-fermion representation. We assume $f^\dagger_{i;a\sigma}$ creates one hole with orbital $a$ and spin $\sigma$ at the site $i$. With the constraint $n_{i;f}=1$, we can recover  the spin-orbital model Hilbert space. Then the spin-orbital model is written as:

\begin{equation}
    H_J=\frac{J}{2}\sum_{i} \sum_{l=1,2,3}  T_{l;aa'} T^\dagger_{l;bb'}(f_{i;a\sigma}^\dagger f_{i;b'\sigma'}) (f^\dagger_{i+a_l;b\sigma'}f_{i+a_l;a'\sigma}) 
\end{equation}
where $J=\frac{4}{U}$ assuming $t_\sigma=1$. We have assumed Einstein summation convention.

At filling $n=1-p$, the low energy physics is described by a $t-J$ model:

\begin{equation}
    H_{t-J}=P H_K P +H_J
\end{equation}
where $P$ is the projection operator to remove the double occupancy.

We will focus on the electron doped side with $p>0$. The other side with hole density $n>1$ may be different in the sense that the additional holes may enter oxygen or Pb atoms, similar to  the hole doped cuprates. We leave it to future to model the hole doped side.

\textbf{Slave boson mean field theory} We use the standard slave boson construction\cite{lee2006doping}: $d_{i;a\sigma}=b_i^\dagger f_{i;a\sigma}$ with the constraint $n_{i;b}+n_{i;f}=1$.  On average $n_{i;f}=1-p$ and $n_{i;b}=p$. Assuming that the slave boson condenses with $\langle b_i \rangle=\sqrt{p}$, we can get the mean field equation for $f$. In the following for convenience we still use  the convention $\psi_{i;\sigma}=(f_{i;1\sigma},f_{i;2\sigma})^T$.

We have the mean field ansatz:
\begin{equation}
    H_M=H_0+H_D
\end{equation}
where

\begin{equation}
    H_0=-\sum_{\sigma}\sum_{i} \sum_{l=1,2,3,4,5,6} \psi_{i;\sigma}^\dagger (p T_l+C_l) \psi_{i+a_l;\sigma}
\end{equation}
The Hermitian condition again constrains that $T_{-l}=T_l^\dagger$ and $C_{-l}=C_l^\dagger$. $C_l$ is a $2\times 2$ complex matrix from decoupling of the super-exchange J term.   Note also $-l$ means the opposite direction of $a_l$.

Meanwhile, we have the pairing term:
\begin{equation}
    H_D=-\sum_{i}\sum_{l=1,2,3,4,5,6} \psi^\dagger_{i;\uparrow}D_l \psi^\dagger_{i+a_l;\downarrow}+H.C.
\end{equation}
We will restrict to spin-singlet pairing, which gives us the constraint that $D_{-l}=D_l^T$.

We have the self consistent equations:

\begin{equation}
    C_l=\frac{J}{2} T_l  \text{Tr} (\chi_l^\dagger T_l^*)
\end{equation}
and
\begin{equation}
   D_l =\frac{J}{4} T_l \Delta_l^T T^*_l
\end{equation}

with the definition:

\begin{align}
    \chi_{l;ab}&=2 \langle \psi^\dagger_{i;a\uparrow} \psi_{i+a_l;b\uparrow} \rangle \notag \\ 
    \Delta_{l;ab}&=2 \langle \psi_{i;a\downarrow} \psi_{i+a_l;b\uparrow} \rangle
\end{align}

\textbf{s wave pairing} We perform the self-consistent calculation at $J/t=0.5$.  We start from initial ansatz with $C_l$ and $D_l$ as random complex matrix.  Nevertheless, we always  reach a time reversal invariant ansatz: $ D_1=\begin{pmatrix}
        0.0121 &-0.0044  \\ 0.0044 &-0.0016 
    \end{pmatrix}$, $ D_2=\begin{pmatrix}
        0.0018 &0.0104 \\ 0.0015 &0.0087 
    \end{pmatrix}$ and   $ D_3=\begin{pmatrix}
        0.0018 &-0.0104  \\ -0.0015 &0.0087
    \end{pmatrix}$ for $p=0.1$.

\begin{figure}
    \centering
    \includegraphics[width=\linewidth]{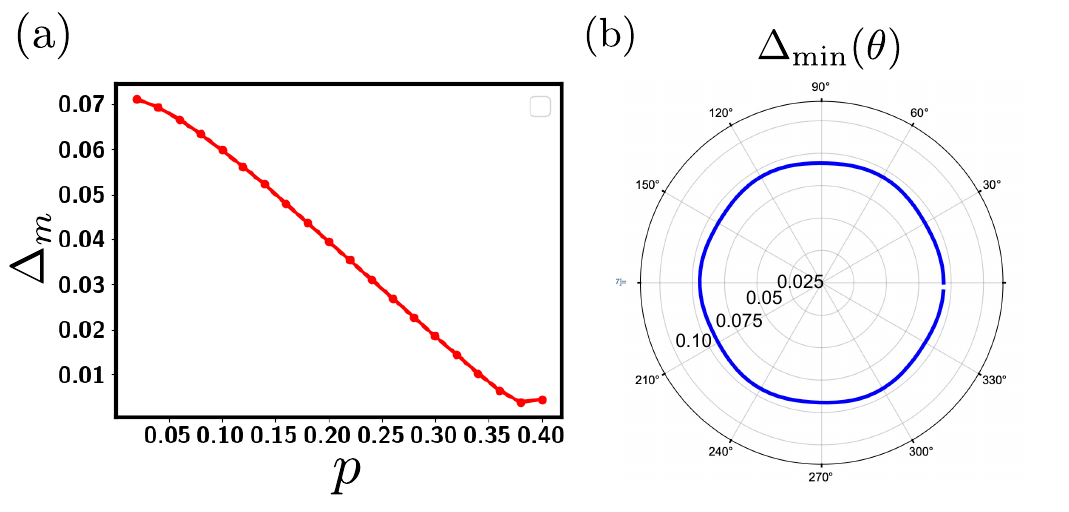}
    \caption{(a) Superconductor gap dependence on doping $p$. $\Delta_m$ is the minimal gap in momentum space. $\Delta_m>0$ suggests that there is no node. (b) At each $\theta$, the radius $r(\theta)$ of the blue line indicates the minimal gap $\Delta_{min}(\theta)$ along this direction. We can see that the gap is non-zero and almost uniform along all directions. Given that the ansatz is time reversal invariant, this is a s-wave pairing. }
    \label{fig:pairing}
\end{figure}

Under $C_3$, we know $\psi \rightarrow U\psi$ with $U=\begin{pmatrix} -\frac{1}{2}& \frac{\sqrt{3}}{2}\\ -\frac{\sqrt{3}}{2} & -\frac{1}{2}\end{pmatrix}$ and then $D_{l} \rightarrow U^\dagger D_l U^*$. We have checked that our ansatz is $C_3$ invariant, like a s-wave pairing.  Similarly it is invariant under $M_y$.  We can project the pairing term $D_{ij}$ to the lower band and get a scalar $\Delta(\mathbf k)$ in momentum space, so the projected Hamiltonian is $\sum_{k} \Delta(\mathbf k) c^\dagger_{\uparrow}(\mathbf k)c^\dagger_{\downarrow}(-\mathbf k)$, where $c_{\sigma}(\mathbf k)$ is the projected operator to the lower band.  $\Delta(\mathbf k)$ is real as guaranteed by the time reversal symmetry: $\psi_{i;a \uparrow} \rightarrow \psi_{i;a \downarrow}, \psi_{i;a\downarrow}\rightarrow -\psi_{i;a\uparrow}$, which must act as $c_{\uparrow}(\mathbf k)\rightarrow c_{\downarrow}(-\mathbf k), c_{\downarrow}(\mathbf k)\rightarrow -c_{\uparrow}(-\mathbf k)$ projected to the lower band.   Meanwhile in the lower band $C_3$ acts trivially and we simply have a requirement $\Delta(C_3 \mathbf k)=\Delta(\mathbf k)$, in agreement with a s-wave pairing.  In Fig.~\ref{fig:pairing} we show that the minimal gap in momentum space is non-zero, suggesting that there is no node.

The s-wave pairing ansatz is robust to variation of $t_2$ and $t_\pi$. We also tried to relax the constraint $D_{-l}=D_l^
T$, but the convergent solution always satisfy this equation, suggesting that the s-wave spin-singlet pairing is the appropriate solution. We also show the dependence of  the pairing gap on doping $p$ in Fig.~\ref{fig:pairing}(a).

\textbf{Discussion}  In the slave boson treatment of the two-orbital t-J model, we find a s-wave pairing and its pairing strength decreases with the doping level $p$, similar to the solution found in the simple t-J model on square lattice. It is known that the real $T_c$ is decided by the phase stiffness at small doping, thus we expect a superconductor dome.  At $p=0.1$, the superconductor gap is $\Delta \approx 0.06 t_\sigma$ at $J/t=0.5$. The DFT calculations suggest that the bandwidth is $W\approx 130$ meV, which  suggests that $t_{\sigma}\approx 20$ meV. Then the gap is only $1.2$ meV and $T_c$ may be only at order of $10$ K. Besides, if $t$ is too small, then $J/t \sim \frac{t}{U}$ can not reach $0.5$ and $T_c$ must be even much smaller.  We note that the $p=0$ insulator may be a charge transfer insulator and there may be other paths to generate a larger $J  $, but $J/t$ seems to be already quite significant.  Even assuming $J/t=0.5$, to get a $T_c$ around 100 K, we need to increase the hopping $t$ by at least one order of magnitude.

Therefore it appears that the model with the value predicted by DFT can not explain the potential high Tc superconductor. New ingredients are needed to enhance the mobility of electron to reach a high temperature superconductor in our model.  One may wonder whether a flat band is beneficial in the phonon driven mechanism. We note that if the hopping is too small, the system is deep inside a strong Mott insulator at $p=0$. Then at finite $p$, the doped electrons will have even reduced hopping due to Mott physics and may tend to localize. It is likely that the Bardeen–Cooper–Schrieffer theory (BCS) theory can not be applied in this case. Due to strong repulsion, a phonon-driven strong superconductor at the small hopping regime does not seem very likely. A larger hopping is probably also needed to reach a high Tc superconductor in the phonon scenario.

So how can we get a higher mobility? We note that the small hopping in this system is largely due to the large Cu-Cu distance and Cu-oxygen distance.  LK-99 system differs from other superconductor materials in the following aspects. In conventional material like cuprates, the doped atoms only provide additional charges and usually do not play any essential roles. The electrons are still moving in the crystal formed by the original atoms. In contrast, in the LK-99 system, one Pb atom per unit cell is replaced with Cu atom. The doped electrons are moving along the array formed by the  Cu atoms. The DFT calculations assume that the Cu atoms just form a periodic crystal, but this assumption is highly questionable. The real chemistry is likely much more complicated.  We conjecture a scenario that the Cu atoms may greatly distort the original crystal and form clusters where the lattice is compressed in a local region and Cu-O and Cu-Cu distances are smaller. The hopping in these clusters can thus be large and locally form s wave pairing according to our calculation.  Then these superconductor islands need to couple together through Josephson tunneling. In the ideal case, there is a global condensation. However, this is not guaranteed. Similar to previous studies of disorder driven superconductor to insulator transition, the system may also be insulating or in the more exotic scenario forms an anomalous metal\cite{kapitulnik2019colloquium}.  Such an anomalous metal has preformed pairs and may explain certain experimental result with diamagnetism, but finite resistivity\cite{lee2023superconductor}.

\textbf{Summary} In conclusion, we propose a two-orbital t-J model to describe the essential physics of LK-99, the candidate material with possible room temperature superconductivity. Our theoretical calculation predicts a s wave pairing. Given that t-J model usually predicts a higher angular momentum pairing, the model is certainly conceptual interesting and worth future theoretical and numerical studies. It is also interesting to explore the possibility that phonon cooperates with the super-exchange $J$ to further enhance the pairing strength. The relevance of the model to LK-99 remains to be investigated in future experimental studies.  To obtain a  superconductor with $T_c\sim 100$K, we need the value of $t$ to be at least one magnitude larger than the predicted values from DFT calculations.  The exact mechanism to generate such a large hopping is not clear now, but may be related to the distortion caused by the Cu atoms. We conjecture that there may be local region which is compressed and thus has a large hopping and a strong s wave pairing according to our model.  This picture suggest that LK-99 may be in the category of granular superconductor. If true, there is likely strong sample dependence with possible superconductor, insulator and even anomalous metal depending on sample details.   We also propose to realize our model and the possible high Tc s-wave superconductor in other materials with active $d_{xz}$ and $d_{yz}$ orbitals.

\textbf{Acknowledgement} YHZ was supported by the
National Science Foundation under Grant No.DMR-2237031. 

%\bibliography{refs}

\begin{thebibliography}{15}%
\makeatletter
\providecommand \@ifxundefined [1]{%
 \@ifx{#1\undefined}
}%
\providecommand \@ifnum [1]{%
 \ifnum #1\expandafter \@firstoftwo
 \else \expandafter \@secondoftwo
 \fi
}%
\providecommand \@ifx [1]{%
 \ifx #1\expandafter \@firstoftwo
 \else \expandafter \@secondoftwo
 \fi
}%
\providecommand \natexlab [1]{#1}%
\providecommand \enquote  [1]{``#1''}%
\providecommand \bibnamefont  [1]{#1}%
\providecommand \bibfnamefont [1]{#1}%
\providecommand \citenamefont [1]{#1}%
\providecommand \href@noop [0]{\@secondoftwo}%
\providecommand \href [0]{\begingroup \@sanitize@url \@href}%
\providecommand \@href[1]{\@@startlink{#1}\@@href}%
\providecommand \@@href[1]{\endgroup#1\@@endlink}%
\providecommand \@sanitize@url [0]{\catcode `\\12\catcode `\$12\catcode
  `\&12\catcode `\#12\catcode `\^12\catcode `\_12\catcode `\%12\relax}%
\providecommand \@@startlink[1]{}%
\providecommand \@@endlink[0]{}%
\providecommand \url  [0]{\begingroup\@sanitize@url \@url }%
\providecommand \@url [1]{\endgroup\@href {#1}{\urlprefix }}%
\providecommand \urlprefix  [0]{URL }%
\providecommand \Eprint [0]{\href }%
\providecommand \doibase [0]{http://dx.doi.org/}%
\providecommand \selectlanguage [0]{\@gobble}%
\providecommand \bibinfo  [0]{\@secondoftwo}%
\providecommand \bibfield  [0]{\@secondoftwo}%
\providecommand \translation [1]{[#1]}%
\providecommand \BibitemOpen [0]{}%
\providecommand \bibitemStop [0]{}%
\providecommand \bibitemNoStop [0]{.\EOS\space}%
\providecommand \EOS [0]{\spacefactor3000\relax}%
\providecommand \BibitemShut  [1]{\csname bibitem#1\endcsname}%
\let\auto@bib@innerbib\@empty
%</preamble>
\bibitem [{\citenamefont {Lee}\ \emph {et~al.}(2023{\natexlab{a}})\citenamefont
  {Lee}, \citenamefont {Kim},\ and\ \citenamefont
  {Kwon}}]{lee2023roomtemperature}%
  \BibitemOpen
  \bibfield  {author} {\bibinfo {author} {\bibfnamefont {S.}~\bibnamefont
  {Lee}}, \bibinfo {author} {\bibfnamefont {J.-H.}\ \bibnamefont {Kim}}, \ and\
  \bibinfo {author} {\bibfnamefont {Y.-W.}\ \bibnamefont {Kwon}},\ }\href@noop
  {} {\enquote {\bibinfo {title} {The first room-temperature ambient-pressure
  superconductor},}\ } (\bibinfo {year} {2023}{\natexlab{a}}),\ \Eprint
  {http://arxiv.org/abs/2307.12008} {arXiv:2307.12008 [cond-mat.supr-con]}
  \BibitemShut {NoStop}%
\bibitem [{\citenamefont {Lee}\ \emph {et~al.}(2023{\natexlab{b}})\citenamefont
  {Lee}, \citenamefont {Kim}, \citenamefont {Kim}, \citenamefont {Im},
  \citenamefont {An},\ and\ \citenamefont {Auh}}]{lee2023superconductor}%
  \BibitemOpen
  \bibfield  {author} {\bibinfo {author} {\bibfnamefont {S.}~\bibnamefont
  {Lee}}, \bibinfo {author} {\bibfnamefont {J.}~\bibnamefont {Kim}}, \bibinfo
  {author} {\bibfnamefont {H.-T.}\ \bibnamefont {Kim}}, \bibinfo {author}
  {\bibfnamefont {S.}~\bibnamefont {Im}}, \bibinfo {author} {\bibfnamefont
  {S.}~\bibnamefont {An}}, \ and\ \bibinfo {author} {\bibfnamefont {K.~H.}\
  \bibnamefont {Auh}},\ }\href@noop {} {\enquote {\bibinfo {title}
  {Superconductor pb$_{10-x}$cu$_x$(po$_4$)$_6$o showing levitation at room
  temperature and atmospheric pressure and mechanism},}\ } (\bibinfo {year}
  {2023}{\natexlab{b}}),\ \Eprint {http://arxiv.org/abs/2307.12037}
  {arXiv:2307.12037 [cond-mat.supr-con]} \BibitemShut {NoStop}%
\bibitem [{\citenamefont {Liu}\ \emph {et~al.}(2023)\citenamefont {Liu},
  \citenamefont {Meng}, \citenamefont {Wang}, \citenamefont {Chen},
  \citenamefont {Duan}, \citenamefont {Zhou}, \citenamefont {Yan},
  \citenamefont {Qin},\ and\ \citenamefont {Liu}}]{liu2023semiconducting}%
  \BibitemOpen
  \bibfield  {author} {\bibinfo {author} {\bibfnamefont {L.}~\bibnamefont
  {Liu}}, \bibinfo {author} {\bibfnamefont {Z.}~\bibnamefont {Meng}}, \bibinfo
  {author} {\bibfnamefont {X.}~\bibnamefont {Wang}}, \bibinfo {author}
  {\bibfnamefont {H.}~\bibnamefont {Chen}}, \bibinfo {author} {\bibfnamefont
  {Z.}~\bibnamefont {Duan}}, \bibinfo {author} {\bibfnamefont {X.}~\bibnamefont
  {Zhou}}, \bibinfo {author} {\bibfnamefont {H.}~\bibnamefont {Yan}}, \bibinfo
  {author} {\bibfnamefont {P.}~\bibnamefont {Qin}}, \ and\ \bibinfo {author}
  {\bibfnamefont {Z.}~\bibnamefont {Liu}},\ }\href@noop {} {\bibfield
  {journal} {\bibinfo  {journal} {arXiv preprint arXiv:2307.16802}\ } (\bibinfo
  {year} {2023})}\BibitemShut {NoStop}%
\bibitem [{\citenamefont {Kumar}\ \emph {et~al.}(2023)\citenamefont {Kumar},
  \citenamefont {Karn},\ and\ \citenamefont {Awana}}]{kumar2023synthesis}%
  \BibitemOpen
  \bibfield  {author} {\bibinfo {author} {\bibfnamefont {K.}~\bibnamefont
  {Kumar}}, \bibinfo {author} {\bibfnamefont {N.}~\bibnamefont {Karn}}, \ and\
  \bibinfo {author} {\bibfnamefont {V.}~\bibnamefont {Awana}},\ }\href@noop {}
  {\bibfield  {journal} {\bibinfo  {journal} {arXiv preprint arXiv:2307.16402}\
  } (\bibinfo {year} {2023})}\BibitemShut {NoStop}%
\bibitem [{\citenamefont {Hou}\ \emph {et~al.}(2023)\citenamefont {Hou},
  \citenamefont {Wei}, \citenamefont {Zhou}, \citenamefont {Sun},\ and\
  \citenamefont {Shi}}]{Shi}%
  \BibitemOpen
  \bibfield  {author} {\bibinfo {author} {\bibfnamefont {Q.}~\bibnamefont
  {Hou}}, \bibinfo {author} {\bibfnamefont {W.}~\bibnamefont {Wei}}, \bibinfo
  {author} {\bibfnamefont {X.}~\bibnamefont {Zhou}}, \bibinfo {author}
  {\bibfnamefont {Y.}~\bibnamefont {Sun}}, \ and\ \bibinfo {author}
  {\bibfnamefont {Z.}~\bibnamefont {Shi}},\ }\href@noop {} {\bibfield
  {journal} {\bibinfo  {journal} {arXiv preprint arXiv:2308.01192}\ } (\bibinfo
  {year} {2023})}\BibitemShut {NoStop}%
\bibitem [{\citenamefont {Wu}\ \emph {et~al.}(2023)\citenamefont {Wu},
  \citenamefont {Yang}, \citenamefont {Xiao},\ and\ \citenamefont
  {Chang}}]{Chang}%
  \BibitemOpen
  \bibfield  {author} {\bibinfo {author} {\bibfnamefont {H.}~\bibnamefont
  {Wu}}, \bibinfo {author} {\bibfnamefont {L.}~\bibnamefont {Yang}}, \bibinfo
  {author} {\bibfnamefont {B.}~\bibnamefont {Xiao}}, \ and\ \bibinfo {author}
  {\bibfnamefont {H.}~\bibnamefont {Chang}},\ }\href@noop {} {\bibfield
  {journal} {\bibinfo  {journal} {arXiv preprint arXiv:2308.01516}\ } (\bibinfo
  {year} {2023})}\BibitemShut {NoStop}%
\bibitem [{\citenamefont {Abramian}\ \emph {et~al.}(2023)\citenamefont
  {Abramian}, \citenamefont {Kuzanyan}, \citenamefont {Nikoghosyan},
  \citenamefont {Teknowijoyo},\ and\ \citenamefont {Gulian}}]{Abramian}%
  \BibitemOpen
  \bibfield  {author} {\bibinfo {author} {\bibfnamefont {P.}~\bibnamefont
  {Abramian}}, \bibinfo {author} {\bibfnamefont {A.}~\bibnamefont {Kuzanyan}},
  \bibinfo {author} {\bibfnamefont {V.}~\bibnamefont {Nikoghosyan}}, \bibinfo
  {author} {\bibfnamefont {S.}~\bibnamefont {Teknowijoyo}}, \ and\ \bibinfo
  {author} {\bibfnamefont {A.}~\bibnamefont {Gulian}},\ }\href@noop {}
  {\bibfield  {journal} {\bibinfo  {journal} {arXiv preprint arXiv:2308.01723}\
  } (\bibinfo {year} {2023})}\BibitemShut {NoStop}%
\bibitem [{\citenamefont {Griffin}(2023)}]{griffin2023origin}%
  \BibitemOpen
  \bibfield  {author} {\bibinfo {author} {\bibfnamefont {S.~M.}\ \bibnamefont
  {Griffin}},\ }\href@noop {} {\bibfield  {journal} {\bibinfo  {journal} {arXiv
  preprint arXiv:2307.16892}\ } (\bibinfo {year} {2023})}\BibitemShut {NoStop}%
\bibitem [{\citenamefont {Kurleto}\ \emph {et~al.}(2023)\citenamefont
  {Kurleto}, \citenamefont {Lany}, \citenamefont {Pashov}, \citenamefont
  {Acharya}, \citenamefont {van Schilfgaarde},\ and\ \citenamefont
  {Dessau}}]{kurleto}%
  \BibitemOpen
  \bibfield  {author} {\bibinfo {author} {\bibfnamefont {R.}~\bibnamefont
  {Kurleto}}, \bibinfo {author} {\bibfnamefont {S.}~\bibnamefont {Lany}},
  \bibinfo {author} {\bibfnamefont {D.}~\bibnamefont {Pashov}}, \bibinfo
  {author} {\bibfnamefont {S.}~\bibnamefont {Acharya}}, \bibinfo {author}
  {\bibfnamefont {M.}~\bibnamefont {van Schilfgaarde}}, \ and\ \bibinfo
  {author} {\bibfnamefont {D.~S.}\ \bibnamefont {Dessau}},\ }\href@noop {}
  {\bibfield  {journal} {\bibinfo  {journal} {arXiv preprint arXiv:2308.00698}\
  } (\bibinfo {year} {2023})}\BibitemShut {NoStop}%
\bibitem [{\citenamefont {Si}\ and\ \citenamefont {Held}(2023)}]{Si}%
  \BibitemOpen
  \bibfield  {author} {\bibinfo {author} {\bibfnamefont {L.}~\bibnamefont
  {Si}}\ and\ \bibinfo {author} {\bibfnamefont {K.}~\bibnamefont {Held}},\
  }\href@noop {} {\bibfield  {journal} {\bibinfo  {journal} {arXiv preprint
  arXiv:2308.00676}\ } (\bibinfo {year} {2023})}\BibitemShut {NoStop}%
\bibitem [{\citenamefont {Tavakol}\ and\ \citenamefont
  {Scaffidi}(2023)}]{Scaffidi}%
  \BibitemOpen
  \bibfield  {author} {\bibinfo {author} {\bibfnamefont {O.}~\bibnamefont
  {Tavakol}}\ and\ \bibinfo {author} {\bibfnamefont {T.}~\bibnamefont
  {Scaffidi}},\ }\href@noop {} {\bibfield  {journal} {\bibinfo  {journal}
  {arXiv preprint arXiv:2308.01315}\ } (\bibinfo {year} {2023})}\BibitemShut
  {NoStop}%
\bibitem [{\citenamefont {Baskaran}(2023)}]{Baskaran}%
  \BibitemOpen
  \bibfield  {author} {\bibinfo {author} {\bibfnamefont {G.}~\bibnamefont
  {Baskaran}},\ }\href@noop {} {\bibfield  {journal} {\bibinfo  {journal}
  {arXiv preprint arXiv:2308.01307}\ } (\bibinfo {year} {2023})}\BibitemShut
  {NoStop}%
\bibitem [{\citenamefont {Lee}\ \emph {et~al.}(2006)\citenamefont {Lee},
  \citenamefont {Nagaosa},\ and\ \citenamefont {Wen}}]{lee2006doping}%
  \BibitemOpen
  \bibfield  {author} {\bibinfo {author} {\bibfnamefont {P.~A.}\ \bibnamefont
  {Lee}}, \bibinfo {author} {\bibfnamefont {N.}~\bibnamefont {Nagaosa}}, \ and\
  \bibinfo {author} {\bibfnamefont {X.-G.}\ \bibnamefont {Wen}},\ }\href@noop
  {} {\bibfield  {journal} {\bibinfo  {journal} {Reviews of modern physics}\
  }\textbf {\bibinfo {volume} {78}},\ \bibinfo {pages} {17} (\bibinfo {year}
  {2006})}\BibitemShut {NoStop}%
\bibitem [{Note1()}]{Note1}%
  \BibitemOpen
  \bibinfo {note} {The pocket looks like a hole pocket in the plot, but because
  we are using the hole picture, this is an electron pocket.}\BibitemShut
  {Stop}%
\bibitem [{\citenamefont {Kapitulnik}\ \emph {et~al.}(2019)\citenamefont
  {Kapitulnik}, \citenamefont {Kivelson},\ and\ \citenamefont
  {Spivak}}]{kapitulnik2019colloquium}%
  \BibitemOpen
  \bibfield  {author} {\bibinfo {author} {\bibfnamefont {A.}~\bibnamefont
  {Kapitulnik}}, \bibinfo {author} {\bibfnamefont {S.~A.}\ \bibnamefont
  {Kivelson}}, \ and\ \bibinfo {author} {\bibfnamefont {B.}~\bibnamefont
  {Spivak}},\ }\href@noop {} {\bibfield  {journal} {\bibinfo  {journal}
  {Reviews of Modern Physics}\ }\textbf {\bibinfo {volume} {91}},\ \bibinfo
  {pages} {011002} (\bibinfo {year} {2019})}\BibitemShut {NoStop}%
\end{thebibliography}

\clearpage

\appendix
\onecolumngrid

% \section{Notations}
% Some important parameters are summarized (we set $a=1$). 
% \begin{itemize}
%     \item Primitive vectors :
%     \begin{eqnarray}
%         \vec{a}_{1}= (1,0), 
%         \quad
%         \vec{a}_{2}=(\frac{1}{2}, \frac{\sqrt{3}}{2}), 
%         \quad 
%         \vec{a}_{3}=(-\frac{1}{2}, \frac{\sqrt{3}}{2}), 
%     \end{eqnarray}
%     \item Reciprocal vectors :
%         \begin{eqnarray}
%         \vec{b}_{1}= 2\pi (\sqrt{3},-1), 
%         \quad
%         \vec{b}_{2}=(0, \frac{4\pi }{\sqrt{3}}), 
%     \end{eqnarray}
% \end{itemize}

\section{Details on Symmetry analysis}
We consider the generic form of tight-binding model with an nearest-neighbor hopping, 
\begin{eqnarray}
    H_{K}&=&
    -\sum_{\sigma=\uparrow,\downarrow}
    \sum_{i,l}  \Psi^\dagger_{i;\sigma}
\bm{T}_{l}\Psi_{i+\vec{a}_{l};\sigma}+H.c, 
\end{eqnarray}
with $2\times 2$ matrix of hopping matrix  $\bm{T}_{l} = \bm{T}_{-l}^{\dagger}$. 
% \yhz{ I don't think you can make the hopping negative, it breaks time reversal symmetry which should be preerved in the system.}
We start with a general form of $T_1$,
\begin{equation}
    T_1=\frac{1}{2}\begin{pmatrix} t_\sigma & t_1-t_2 \\ t_1+t_2 & t_\pi \end{pmatrix}
\end{equation}

One can find that the mirror symmetry ($\mathcal{M}_{y}$) imposes $t_{1}=0$, since  $T_{-1}=U_{M_y}^\dagger T_1 U_{M_y}$ while $T_{-1}=T_1^\dagger$. We have $U_{M_y}=\begin{pmatrix}-1 & 0 \\ 0 & 1\end{pmatrix}$. It is easy to see that we need $t_1=0$.

Hence we introduce a general tight-binding model with three free variables $t_{\sigma},t_{\pi},t_{2}$. From $T_1$ we can generate $T_2$ and $T_3$ by applying the $C_3$ transformation: $T_3=U^\dagger_{C_3}T_1 U_{C_3}$ and $T_{-2}=T_2^\dagger=U^\dagger_{C_3}T_3 U_{C_3}$. The final tight binding model is constrained to be:
\begin{eqnarray}
    H_{K} &=&
    -\frac{1}{2}\sum_{\sigma }
    \sum_{i,l}
    \Psi_{i}^{\dagger}
    \left(
    \begin{array}{cc}
         t_{\sigma}& - t_{2} \\
       t_{2}  & t_{\pi}
    \end{array}
    \right)
    \Psi_{i+\vec{a}_{1}}
    +H.c.\\
    &&+ 
    \Psi_{i}^{\dagger}
    \left(
    \begin{array}{cc}
\frac {1} {4} (t_\sigma + 3 t_ {\pi}) & 
 t_ {2} + \frac {\sqrt {3}} {4} (t_\sigma- t_\pi) \\
-t_ {2} + \frac {\sqrt {3}} {4} (t_\sigma- t_\pi)
       & \frac {1} {4} (3t_\sigma +  t_ {\pi}) 
    \end{array}
    \right)
    \Psi_{i+\vec{a}_{2}}
    +H.c.\\
    &&+ \Psi_{i}^{\dagger}
    \left(
    \begin{array}{cc}
\frac {1} {4} (t_\sigma + 3 t_ {\pi}) & - 
 t_{2} -\frac {\sqrt {3}} {4} (t_\sigma- t_\pi) \\
t_{2} - \frac {\sqrt {3}} {4} (t_\sigma- t_\pi)
       & \frac {1} {4} (3t_\sigma + t_ {\pi}) 
    \end{array}
    \right)
    \Psi_{i+\vec{a}_{3}}
    +H.c.\\
    &=& 
    \sum_{\sigma} 
\Psi^{\dagger}_{k;\sigma}
    \left(
\begin{array}{cc}
h_{11}(k)  & h_{12}(k)  \\
h_{21}(k)   & h_{22}(k) 
\end{array}\right)
\Psi_{k;\sigma}
    \end{eqnarray}
with
\begin{eqnarray}
h_{ab}(k) &=&
-\sum_{l} \left[
(t_{\sigma} \phi^{a}_{\sigma, l}\phi^{b}_{\sigma, l} 
+t_{\pi} \sum_{l} \phi^{a}_{\pi, l}\phi^{b}_{\pi, l})
\cos(k\cdot \vec{a}_{l})
+t_{2}
\sigma^{ab}_{y}
s^{l}
\sin(k\cdot \vec{a}_{l})
\right]
\end{eqnarray}
% and,
with $ \phi_{\sigma, l}= \vec{a}_{l}/a$, and 
\begin{eqnarray}
   && \phi_{\sigma, 1}=(1,0), 
    \quad \phi_{\sigma, 2}=(\frac{1}{2},\frac{\sqrt{3}}{2}),
    \quad\phi_{\sigma, 3}=(-\frac{1}{2},\frac{\sqrt{3}}{2}), \\
   && \phi_{\pi, 1}=(0,1), 
    \quad \phi_{\pi, 2}=(-\frac{\sqrt{3}}{2},\frac{1}{2}),
    \quad\phi_{\pi, 3}=(-\frac{\sqrt{3}}{2},-\frac{1}{2}).
\end{eqnarray}
with $s_{1} = -s_{2}=s_{3} =1$.

\section{Derivation of the self consistent equations.}

\subsection{mean field Hamiltonian}

We first write down the most general mean field ansatz.  We assume translation symmetry and spin rotation symmetry( only spin-singlet pairing). But allow general other ansatz.

We introduce the notation that $\psi_{i;\sigma}=(f_{i;a\sigma},f_{i;b\sigma})^T$.  Ourm ean field can be written as:

\begin{equation}
    H_M=H_0+H_D
\end{equation}
where

\begin{equation}
    H_0=-\sum_{i,\sigma} \sum_{l=1,2,3,4,5,6} 
    \psi_{i;\sigma}^\dagger (p T_l+C_l) \psi_{i+a_l;\sigma}
\end{equation}
We always have constraint that $T_{-l}=T_l^\dagger$ and $C_{-l}=C_l^\dagger$. $C_l$ is a $2\times 2$ complex matrix from decoupling of the super-exchange J term.

Meanwhile, we have the pairing term:

\begin{equation}
    H_D=-\sum_{i}\sum_{l=1,2,3,4,5,6} \psi^\dagger_{i;\uparrow}D_l \psi^*_{i+a_l;\downarrow}+H.C.
\end{equation}

We have the constraint that $D_{-l}=D_l^T$ for spin-singlet pairing.

Here $D_l$ is a $2\times 2$ matrix decoupled from the J term.

\subsection{Self consistent equations}

Before we continue, we define the following two $2\times 2$ complex matrix:

\begin{align}
    \chi_{l;ab}&=2 \langle \psi^\dagger_{i;a\uparrow} \psi_{i+a_l;b\uparrow} \rangle \notag \\ 
    \Delta_{l;ab}&=2 \langle \psi_{i;a\downarrow} \psi_{i+a_l;b\uparrow} \rangle
\end{align}

We fix the overall hopping term energy scale to be $1$. Then we define $J=\frac{4}{U}$ to match the usual definition. From the second order perturbation, we can write the $J$ term as:

\begin{equation}
    H_J=-\frac{2}{U}\sum_{i} \sum_{l=1,2,3} \psi_{i;a\sigma}^\dagger T_{l;aa'} \psi_{i+a_l;a'\sigma} \psi^\dagger_{i+a_l;b\sigma'}T^\dagger_{l;bb'}\psi_{i;b'\sigma'} 
\end{equation}

Then we can get the following self consistent equation for the hopping term:

\begin{equation}
    C_{l;a a'}=T_{l;a a'} 
    \frac{J}{2} 2 \sum_{b,b'}
    \langle \psi^\dagger_{i+a_l;b\uparrow} T^\dagger_{l;bb'}\psi_{i;b'\uparrow} \rangle
\end{equation}

In matrix notation, it is basically:

\begin{equation}
    C_l=\frac{J}{2} T_l  \sum_{b,b'}\chi^\dagger_{l;bb'}
    T^*_{l;b'b}
=\frac{J}{2} T_l  \mathrm{Tr}
\left(
\chi^\dagger_{l}
    T^*_{l}    \right)
\end{equation}

For pairing, we can get:

\begin{equation}
    D_{l;ab}=\frac{J}{4} \sum_{a'b'} T_{l;aa'}T^\dagger_{l;bb'}\Delta_{l;b'a'}
\end{equation}

which is equivalent to:

\begin{equation}
   D_l =\frac{J}{4} T_l \Delta_l^T T^*_l
\end{equation}

\end{document}